\documentclass[preprint2]{aastex}

\input epsf

\begin{document}

\title{Cooling Rates of Molecular Clouds Based on Numerical MHD Turbulence 
and non-LTE Radiative Transfer}

\author{
Mika Juvela 
\footnote{mjuvela@astro.helsinki.fi}}
\affil{Helsinki University Observatory, T\"ahtitorninm\"aki, P.O.Box 14,
SF-00014 University of Helsinki, Finland}
\author{Paolo Padoan 
\footnote{ppadoan@cfa.harvard.edu}}
\affil{Harvard University, Department of Astronomy, 60 Garden Street, Cambridge, MA 02138}
\and
\author{\AA ke Nordlund}
\affil{Copenhagen Astronomical Observatory, and Theoretical Astrophysics Center, 
2100 Copenhagen, Denmark}

\begin{abstract}

We have computed line emission cooling rates for the main cooling species in
models of interstellar molecular clouds. The models are based on numerical
simulations of super--sonic magneto--hydrodynamic (MHD) turbulence. 
Non-LTE radiative transfer calculations have been performed to properly
account for the complex density and velocity structures in the MHD
simulations. 

Three models are used. Two of the models are based on MHD simulations 
with different magnetic field strength (one model is super--Alfv\'{e}nic,
while the other has equipartition of magnetic and kinetic energy). 
The third model includes the computation of self-gravity (in the 
super--Alfv\'{e}nic regime of turbulence). The density and velocity 
fields in the simulations are determined self--consistently by the 
dynamics of super--sonic turbulence. The models are intended to
represent molecular clouds with linear size $L\approx 6$\,pc and mean
density $\langle n \rangle \approx 300$\,cm$^{-3}$, with the density exceeding
10$^4$\,cm$^{-3}$ in the densest cores.

We present $^{12}$CO, $^{13}$CO, C$^{18}$O, O$_2$, O{\sc I}, C{\sc I} and
H$_2$O cooling rates in isothermal clouds with kinetic temperatures 10--80\,K.
Analytical approximations are derived for the cooling rates.

The inhomogeneity of the models reduces photon trapping and enhances the
cooling in the densest parts of the clouds.
Compared with earlier models the cooling rates are less affected by optical
depth effects. The main effects come, however, from the density variation
since cooling efficiency increases with density. This is very important for
the cooling of the clouds as a whole since most cooling is provided by gas
with density above the average.

\end{abstract}


\section{Introduction}

Line emission by molecules and atomic species is the most important cooling
process in interstellar clouds \citep{goldsmith78}. The
temperatures of molecular clouds are determined by the balance between
radiative cooling and various heating mechanisms like cosmic ray heating,
formation of H$_2$ molecules, photo-electric heating due to external radiation
\citep{dejong77} and ambipolar drift (Padoan, Zweibel \& Nordlund
\citeyear{Padoan+2000ad}).

In order to derive the radiative cooling rate for a given species one must
first know the excitation conditions in the cloud. However, the cooling rates
also depend on the detailed structure of the cloud and the true net flow of
energy can only be obtained by solving the full radiative transfer problem. So
far the calculations have been based on homogeneous models in different
geometries or spherically symmetric clouds with smooth density distribution.

Because of its large abundance the CO molecules is usually the most important
coolant of molecular gas. \citet{goldreich74} studied the
CO cooling and made comparisons between the CO cooling rates and the heating
associated with cloud collapse. The radiative transfer calculations were based
on the Sobolev, or large velocity gradient (LVG), approximation.
\citet{goldsmith78} computed cooling rates due to a number of
molecules and atomic species using the LVG method applied to spherical,
homogeneous and isothermal model clouds. They covered a wide density range up
to 10$^7$\,cm$^{-3}$ and a temperatures up to 60\,K. Although $^{12}$CO
dominates the cooling of low density gas, the rare CO isotopes, C{\rm I} and
O$_2$ were found to contribute a large fraction of the total cooling rates. At
even higher densities the cooling rates are determined by a large number of
species including H$_2$O, various hydrides and molecular ions. Goldsmith
\& Langer considered several heating mechanisms and derived equilibrium
temperatures for typical clouds.

\citet{neufeld95} continued this work by studying radiative
cooling rates of dense molecular clouds, $n=$ 10$^3$-10$^{10}$\,cm$^{-3}$ using
updated molecular data. The molecular gas was assumed to be fully shielded
from external ionizing radiation. Neufeld et al.\ used chemical models to
compute the steady state molecular abundances that were used as the basis of
the cooling rate calculations. Level populations of the main cooling species
were solved using the escape probability formalism \citep{neufeld93}. Strictly
speaking the calculations assume plane-parallel geometry with strong velocity
gradient but the results can be applied also to different geometries. That
paper concentrated on the study of isothermal, spherical models.

In the present work we re--examine the cooling efficiency of the main cooling
agents in molecular clouds at temperatures T$_{\rm kin}<$100\,K. Compared with
the earlier studies there are two main improvements: i) The model clouds
are no longer assumed to be homogeneous and, more importantly, the density and
velocity structures are the result of realistic magnetohydrodynamical
simulations; ii) The non-LTE radiative transfer problem is solved exactly with
Monte Carlo methods. The solution takes fully into account the inhomogeneous
density and velocity fields of the models.

\section{The Cloud Models} \label{sect:model}

The numerical models used in this work are based on the results of numerical
simulations of highly super--sonic magneto--hydrodynamic (MHD) turbulence, run
on a 128$^3$ computational mesh, with periodic boundary conditions. 

As in our previous work, the initial density and magnetic fields are uniform.
%%AA  The initial velocity is in-consequential when we are forcing --
%%AA  better leave it out, to avoid confusion.
We apply an external random force, to drive the turbulence at a roughly
constant rms Mach number of the flow. The force is generated in Fourier space,
with power only at small wave numbers ($1 \leq k \leq 2$). The isothermal
equation of state is used. A description of the numerical code used to solve
the MHD equations may be found in Padoan \& Nordlund
(\citeyear{Padoan+Nordlund98MHD} and references therein).

In order to scale the models to physical units, we use the following empirical
Larson type relations, as in our previous works:
\begin{equation}
{\cal M}_s=4.0\left(\frac{L}{1pc}\right)^{0.5},
\end{equation} 
where ${\cal M}_s$ is the rms sonic Mach number of the flow (the rms flow
velocity divided by the sound speed), and a temperature $T=10$~K is assumed,
and
\begin{equation}
\langle n \rangle=2.0\times10^3\left(\frac{L}{1pc}\right)^{-1},
\end{equation} 
where the gas density $n$ is expressed in cm$^{-3}$. The rms sonic Mach number
is an input parameter of the numerical simulations, and can be used to scale
them to physical units. The rms Alfv\'{e}nic Mach number of the flow ${\cal
M}_a$ is also an input parameter of the numerical simulations. ${\cal M}_a$ is
defined as the ratio of the rms flow velocity and the Alfv\'{e}n velocity,
$(\langle B^2\rangle/4\pi\langle\rho\rangle)^{\frac{1}{2}}$.
%%AA  Is this the Alfven speed actually used?  It should be.  In any case it
%%AA  needs to be defined.
It determines the magnetic field strength, once the sonic rms Mach number,
${\cal M}_s$, is fixed. We refer to the turbulent flow as super--Alfv\'enic
when ${\cal M}_a > 1$, while by equipartition turbulence we mean ${\cal
M}_a\approx 1$.

In this work we use three models. They are all highly super--sonic, with
${\cal M}_s \sim 10$. Models $B$ and $C$ are super--Alfv\'{e}nic, with ${\cal
M}_a \sim 10$, while model $A$ has rough equipartition of magnetic and kinetic
energy of turbulence, with ${\cal M}_a \sim 1$. Models $A$ and $B$ neglect the
effect of self--gravity, which is instead included in model $C$.

The physical unit of velocity in the code is the isothermal speed of sound,
$C_s$, and the physical unit of the magnetic field is
$C_s(4\pi\langle\rho\rangle)^{\frac{1}{2}}$ (cgs). Assuming a kinetic
temperature of $T_{\rm kin}$=10\,K and a mean density of 320\,cm$^{-3}$ the mean
field strength is $47.0$~$\mu$G in model $A$, $2.2$~$\mu$G in model $B$ and
$2.6$~$\mu$G in model $C$. At $T_{\rm kin}$=10\,K the rms velocity is
approximately 2.1~km/s and the linear size $L=6.3$~pc in all three models. The
turbulent velocity inside each computational cell is estimated as the rms
velocity between neighboring cells. The macroscopic velocities and the thermal
line widths are scaled according to the assumed temperature.

\begin{figure}[!th]
\epsscale{1.0}
\plotone{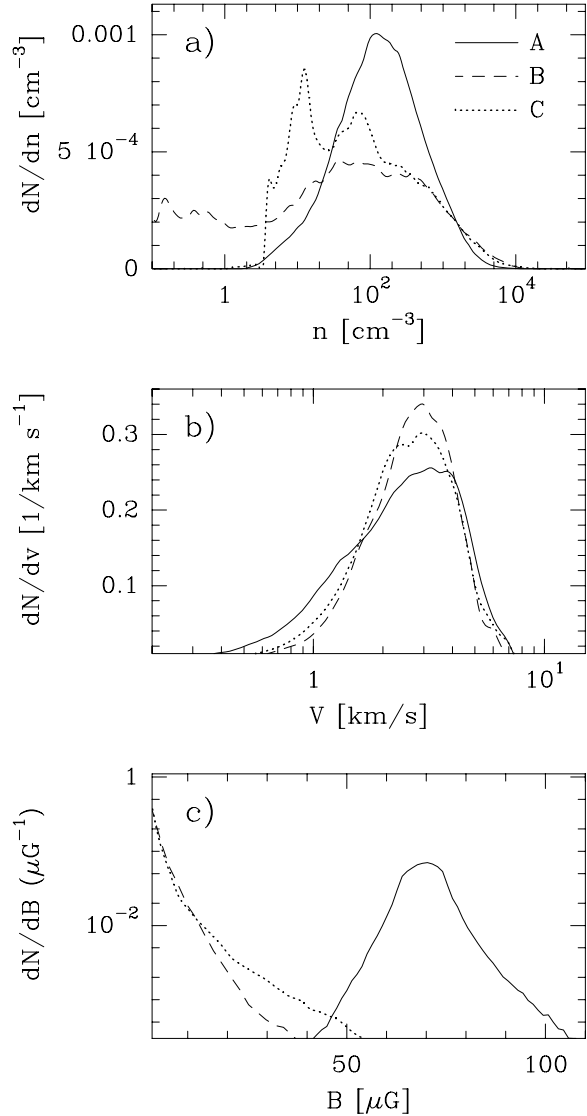}
\caption[]{
Number of cells in the three model clouds ($A$, $B$ and $C$) as a function of
density (a) velocity (b) and magnetic field strength (c). The plots assume a
mean density of 320\,cm$^{-3}$ and a temperature of $T_{\rm kin}$=20\,K. }
\label{fig:distributions}
\end{figure}

In Figure~\ref{fig:distributions} we show the distribution of cell densities,
velocities and magnetic field strength for the three models. Model $B$ has the
widest range of densities and the largest fraction of dense cells,
$n\sim$10$^3$\,cm$^{-3}$ (top panel of Figure~\ref{fig:distributions}). In $C$
the density distribution is skewed towards lower densities. Model $A$ is
the least inhomogeneous one, but the density contrast is still larger than three
orders of magnitude. As can be seen from the middle panel of
Figure~\ref{fig:distributions}, differences in the velocity distribution
between different models are insignificant. The magnetic field strengths are
shown in the bottom panel of Figure~\ref{fig:distributions}, with physical
values obtained for the mean density $n=320$~cm$^{-3}$ and the kinetic
temperature $T_{\rm kin}=20$~K.

\section{The Calculation of the Cooling Rate} \label{sect:rate}

The cooling rates are calculated individually for each cell in the model
clouds and for several cooling species. The original MHD simulations were
performed on a 128$^3$ grid but because of the high computational burden of
the radiative transfer calculations most of these calculations were carried
out with model clouds re-sampled into a 90$^3$ cell grid. The discretization
introduces some small smoothing of the high density peaks but is otherwise not
expected to affect the derived cooling rates.

The collisional coefficients for CO were taken from 
\citet{flower85}, C{\sc I}-H$_2$ rates are from \citet{schroder91}
and O{\sc I}-H$_2$ rates from \citet{jacquet92}. The rate coefficients for
collisions O$_2$-H$_2$ and O$_2$-He were provided by P. Bergman
(\citeyear{bergman95}; private communication).

\subsection{Fractional Abundances of the Cooling Species}

The cooling rates are calculated for the species $^{12}$CO, $^{13}$CO,
C$^{18}$O, O$_2$, C{\sc I} and O{\sc I}. According to 
\citet{neufeld93} these are the most important cooling species in the present
parameter region, that is at densities below $\sim10^4$\,cm$^{-3}$ and at
temperatures $T_{\rm kin}<$100\,K. Cooling rates due to H$_2$O are calculated
only at T$_{\rm kin}$=60\,K.

For the CO species we assume fractional abundances
[$^{12}$CO]/[H$_2$]=$5\cdot10^{-5}$, [$^{13}$CO]/[H$_2$]=$1\cdot10^{-6}$ and
[C$^{18}$O]/[H$_2$]=$1\cdot10^{-7}$. These values are similar to those adopted
by \citet{goldsmith78}. The CO abundance is lower than predicted by standard
chemical models \citep{millar97, lee96} but consistent with observation
(e.g. Ohishi et al.
\citeyear{ohishi92}) which also show a significant variations between clouds
\citep{harjunpaa}. Depending e.g. on the radiation field and the gas temperature, chemical
fractionation can lead to abundance variations (e.g. \citealt{warin96}) but we
shall assume constant abundances throughout the clouds. This is probably a
good approximation since the inhomogeneous density distribution reduces
differences in the radiation field between inner and outer parts of the cloud
\citep{boisse90, spaans96} and we do consider only isothermal models.

The abundances of O$_2$ and O{\sc I} are not well known. Recent results from
the ISO and SWAS satellites show that the oxygen abundances in molecular
clouds have been previously overestimated and probable values are below
[O$_2$]/[H$_2$]$\sim 10^{-6}$ \citep{bergin00, goldsmith00}. On the other
hand, \citet{caux99} found L1689N to be rich in O{\sc I}
indicating that the abundance of O{\sc I} can be as high as [O{\sc
I}]/[H$_2$]$\ga 10^{-5}$. The predictions of chemical models have usually been
closer to 10$^{-4}$ for both [O{\sc I}]/[H$_2$] and [O$_2$]/[H$_2$] (e.g. 
\citealt{lee96}). Above 100\,K the abundances are also very sensitive to
the assumed temperature but in the temperature range considered in the present
work the abundances are essentially independent of temperature 
\citep{neufeld95}. We use an abundance of $10^{-5}$ for both O{\sc I} and
O$_2$. In view of the recent observational results the O{\sc I} abundance is
adequate but the O$_2$ abundance might be too high by more than one order of
magnitude. However, a similar value is used by
\citet{goldsmith78} and \citet{neufeld95}, which makes
comparison with their results easier.
%%% ### @
For C{\sc I} a fractional abundance value of 10$^{-6}$ will be used.

For water a fractional abundance of 1.0$\cdot 10^{-6}$ is assumed.
Calculations are carried out separately for ortho- and para-water with ortho
to para ratio 1:3. The total abundance is similar to the values used by
\citet{goldsmith78} and \citet{neufeld95}. However, observations of quiescent
gas in molecular cloud cores, Orion and M17SW \citep{snell00a, snell00b,
snell00c} have indicated much lower abundances [H$_2$O]/[H$_2$]$\sim10^{-8}$.
Similar values have been reported by \citet{ashby00}. The lower values would
mean that water is unimportant for the cooling of the clouds considered here.
%% ##### @
Locally H$_2$O can be very efficient coolant since in outflows and hot cores
its abundance can be enhanced up to [H$_2$O]/[H$_2$]$\sim10^{-4}$
\citep{snell00c} and the importance of H$_2$O increases with temperature. Our
models represent, however, relatively cold and quiescent clouds. The H$_2$O
cooling rates are therefore computed only for $T_{\rm kin}$=60\,K and with
relative abundance [H$_2$O]/[H$_2$]=1.0$\cdot 10^{-6}$ the results can be
taken as upper limits for the actual H$_2$O cooling.
%% #####

Since we will examine only isothermal models, the possible temperature
dependence of the abundances affects only the comparison between models with
different $T_{\rm kin}$. At these low temperatures the temperature dependence
is, however, weak. According to models (e.g. \citealt{lee96}) there
can be a significant dependence on the gas density even when photo-processes
due to external radiation field are not considered. The effect is especially
clear for carbon. In the standard model presented by \citet{lee96}
the carbon abundance increases by a factor $\sim$10 as the gas density
decreases from 10$^4$\,cm$^{-3}$ to 10$^3$\,cm$^{-3}$. This would reduce the
spatial variation in the carbon emission. We have not included these abundance
variations in our models. The steady state abundances are not necessarily
valid for the turbulent medium of the MHD models where density variations are
caused by moving shock fronts and the predicted abundances must be treated
with some caution.

\subsection{Energy Levels Used in the Calculations}

For practical reasons the number of energy levels that can be included in the
calculations is limited. The number of levels needed for an accurate estimate
of the cooling rate depends mainly on the excitation as all significantly
populated levels must be considered. The relative importance of the
transitions is affected also by optical depth, as optically thick transitions
contribute less to the total cooling rate.

%%% ### @
In the case of $^{12}$CO we find that at low temperatures, $T_{\rm
kin}\le20$\,K, it is sufficient to include levels up to $J=11$. 
Higher transitions are more important at higher densities. However, at 20\,K,
the cooling rate of the $J=8-7$ transition is approximately two orders of
magnitude below that of the $J=2-1$ transition, even for densities
$n\ga$10$^3$\,cm$^{-3}$. The total contribution from higher transitions is
insignificant.

\begin{figure}[!th]
\plotone{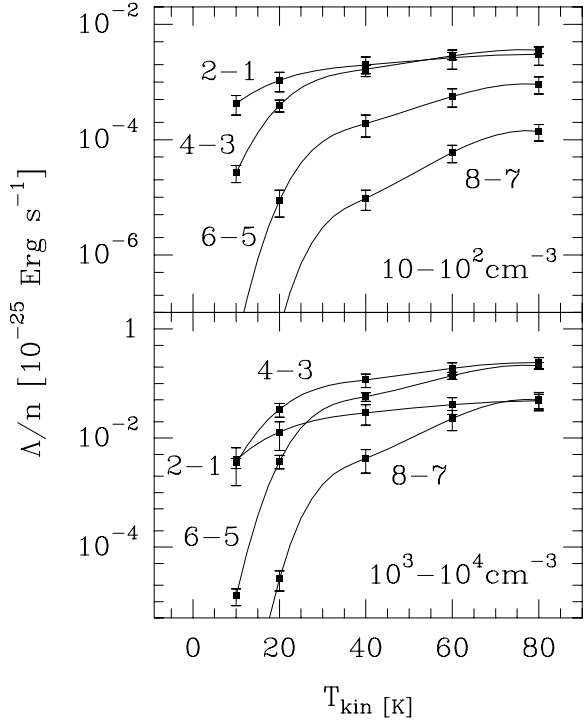}
\caption[]{%
The cooling rates, $\Lambda/n$, due to selected $^{12}$CO transitions
($J=$2-1, 4-3, 6-5 and 8-7) as a function of the kinetic temperature. The
rates are averaged over cells in the density ranges 10$^3$-10$^4$\,cm$^{-3}$
(upper panel) and 10-10$^2$\,cm$^{-3}$ (lower panel). The errorbars correspond to the
dispersions within the given density ranges.}
\label{fig:cbt_n_T}
\end{figure}

At higher temperatures, $T_{\rm kin}>20$\,K, we include the 15 lowest energy
levels of $^{12}$CO. In Figure~\ref{fig:cbt_n_T} the cooling rate in model $B$
is plotted as a function of the kinetic temperature. At $T_{\rm kin}\ge 60$\,K
and at densities $n\sim$10$^4$\,cm$^{-3}$ the cooling rate from the $J=8-7$
transition is roughly equal to the rate from the $J=2-1$ transition. However,
the population as a function of $J$ decreases rapidly
and the contribution of levels close to $J=14$ is insignificant. All models
have similar ranges of density and the same number of levels is used also for
the other two model clouds ($A$ and $C$).

Due to small optical depths the number of populated $^{13}$CO and C$^{18}$O
levels is lower than in $^{12}$CO. For these species we include 10 levels at
$T_{\rm kin}\le$20\,K and 12 at higher temperatures. The number of O$_2$
levels used in the calculations was the same as for CO but because of the fine
structure there are now 22 transitions between the first 15 energy levels. For
both C{\sc I} and O{\sc I}, 3 energy levels were used, with only 2 transitions
between them. The C{\rm I} lines are $^3$P$_1$-$^3$P$_0$ at 610\,$\mu$m and
$^3$P$_2$-$^3$P$_0$ at 230\,$\mu$m and the corresponding O{\sc I} lines are at
145\,$\mu$m and 63\,$\mu$m. Cooling rates due to H$_2$O were calculated only
at 60\,K and only 11 energy levels (20 transitions) were
included. This is adequate due to the relatively low densities and
temperatures. The highest included levels are more than $400$\,K above the
ground state and are not significantly populated.

The inhomogeneous cloud structure of the MHD models increases the photon
escape probability in otherwise optically thick transitions. The cooling rate
from lower transitions is increased while the population of the higher levels
is decreased.

\section{The Radiative Transfer Method}

The model cloud is divided into 90$^3$ cells. Each cell is assumed to be
homogeneous and is characterized by one value of density, intrinsic linewidth
and macroscopic velocity.
%%% ### @
Density and velocity are obtained directly from MHD simulations once the
results are scaled to physical units. The intrinsic linewidth is the sum of
thermal line broadening and the Doppler broadening caused by turbulent motions
inside the cell. The latter is estimated as the rms velocity difference
between neighboring cells. The components are of the same order but usually
the turbulent line broadening is the larger of the two.
%%% ###

The radiative transfer problem is solved with a Monte Carlo method (method B
in \citealt{juvela97}). The radiation field is simulated by a large number of
photon packages going through the cloud. These represent both photons entering
the cloud from the background and photons emitted within the cloud. As a
photon package goes through a cell the number of photons absorbed within that
cell is removed from the package. At the same time the number of upward
transitions induced by these photons is stored in counters. There is a
separate counter for each cell and each simulated transition. After the
simulation of the radiation field the counters are used to obtain new
estimates of the level populations. The whole process is repeated until the
level populations converge (the relative change from an iteration to the next
is less than 10$^{-4}$).
The core saturation method was used to speed the calculations of
optically thick species \citep{hartstein, alma}.

The cooling rates of the cells are also calculated with a Monte Carlo method,
using the previously obtained level populations. During the computation the net
flux of photons is counted for each cell and this is transformed into cooling
rates in units of erg\,s$^{-1}$\,cm$^{-3}$. In order to study the relative
importance of the transitions the net flux is counted for each transition
separately.

\begin{figure}[!th]
\plotone{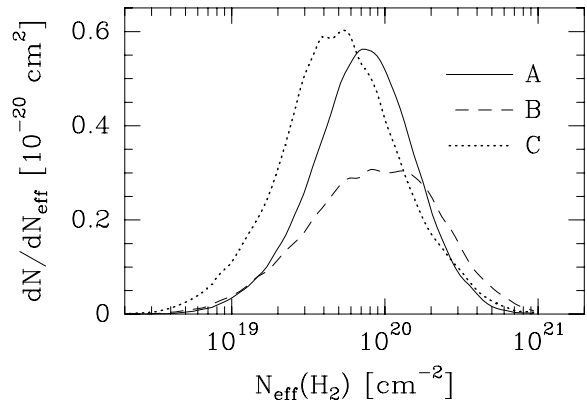}
\caption[]{%
Distributions of effective hydrogen column density, N$_{\rm eff}$, 
in the three model clouds. The average density is 320\,cm$^{-3}$ 
and the linear size 6.3~pc in all models.}
\label{fig:nn_distribution}
\end{figure}

\section{Results} \label{sect:result}

\subsection{Cooling Rates} \label{sect:rates}

The local cooling rate depends mainly on three parameters: the local density,
kinetic temperature and effective column density, or effective optical depth.
The optical depth determines the photon escape probability that in our case is
strongly affected by the inhomogeneity of the clouds. The optical depth seen
by a cell varies strongly depending on the line of sight. This is true
throughout the cloud, not only close to the cloud surface. The effective
optical depth depends on both the density and the velocity distributions since
the velocity dispersion is always large compared with the thermal linewidth.
The velocity dispersion in the models is approximately 3.0~km/s, assuming
a sound speed 0.3~km/s.

\begin{figure}[!th]
\plotone{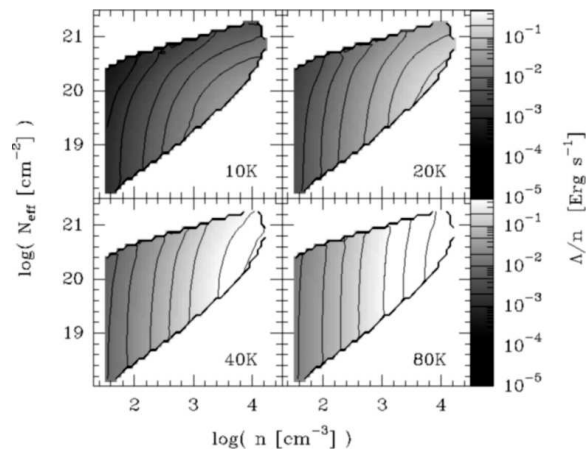}
\caption[]{%
$^{12}$CO cooling rates $\Lambda/n$ as a function of density, 
$n$, and effective column density, $N_{\rm eff}$, in model $B$.}
\label{fig:nnplane}
\end{figure}

In order to quantify these effects, we have calculated the effective column
density, $N_{\rm eff}$, seen by each cell in the three models. We define
this as 
\begin{equation}
N_{\rm eff} \,\, = \,\, \langle \,\,
\left( \int N(v) \phi(v) dv \right) ^{-1}
\,\, \rangle ^{-1}.  
\label{eq:neff}
\end{equation}
$N(v)$ is the column density along one line of sight from the edge of the
cloud to the cell and $\phi(v)$ is the local absorption profile, both given in
units of velocity. The integral is proportional to the column density seen by
the cell towards one direction and the averaging is done over all directions.
$N_{\rm eff}$ is calculated as a by-product of the radiative transfer
calculations. The cooling rate depends on $N_{\rm eff}$ through its effect on
the effective optical depth seen by the cells i.e. the escape probability. The
optical depth depends in a complicated way on the excitation in other parts of
the cloud. The effective column density is, on the other hand, a parameter of
the cloud itself and describes the effects of the density and velocity fields
independently of the studied molecule. For these reason the effective column
density will be used as a substitute for the effective optical depth.

The distributions of $N_{\rm eff}$ seen by individual cells are shown in
Figure~\ref{fig:nn_distribution} for the three models.
%%% ### @
$N_{\rm eff}$ is smallest on the surface of the cloud and increases towards
the centre depending, however, on the actual density and velocity fields. In
microturbulent case with gaussian line profiles we would get $N_{\rm
eff}=N/\sqrt{4\pi\,\sigma^2}$. Here $N$ is approximately equal to the average
column density between the cell and the cloud boundary. In our case we have
$<n>=320$\,cm$^{-3}$ and $L/2$=3.1\,pc. When total velocity dispersion, some
3\,km\,s$^{-1}$ at 20\,K, is used for $\sigma$ we obtain an effective column
density of $\sim$3$\cdot$10$^{20}$cm$^{-2}$\,km$^{-1}$\,s in the cloud centre.
In MHD simulations the turbulence is, however, not random and more importantly
the density distribution is not constant. Higher values, up to $N_{\rm
eff}\sim 10^{21}$cm$^{-2}$\,km$^{-1}$\,s, are therefore reached in dense and
velocity coherent regions.
%%% ###

\begin{figure}[!th]
\epsscale{1.0}
\plotone{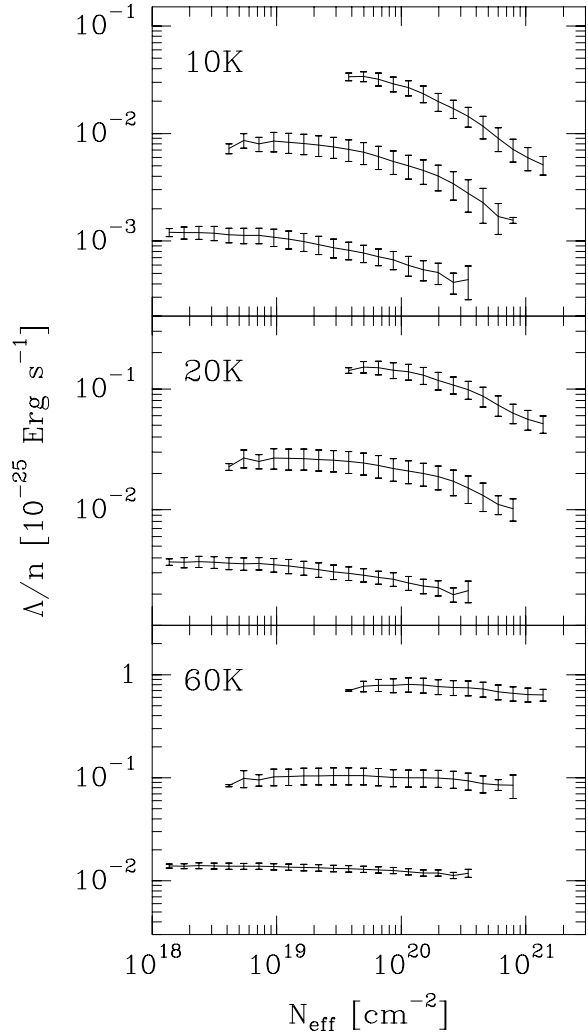}
\caption[]{%
Cooling rate $\Lambda({\rm CO})/n$ in model $B$ as a function of the effective
column density $N_{\rm eff}$, in clouds with kinetic temperature T$_{\rm
kin}$=10, 20 or 60\,K. Each panel shows the average cooling rates in three
density intervals, 2.2-4.2$\cdot$10$^1$\,cm$^{-3}$ (lowest curves),
2.2$\cdot$10$^2$-4.2$\cdot$10$^2$\,cm$^{-3}$ and
2.2$\cdot$10$^3$-4.2$\cdot$10$^3$\,cm$^{-3}$ (highest curves). The errorbars
reflect the variation of $\Lambda({\rm CO})/n$ in the given density
intervals.}
\label{fig:coolcolden}
\end{figure}

Figure~\ref{fig:nnplane} summarizes the CO cooling rate for model $B$ as a
function of density, $n$, effective column density, $N_{\rm eff}$, and
temperature. In Figure~\ref{fig:coolcolden} the cooling rate $\Lambda({\rm
CO})/n$ in model $B$ is plotted as a function of the effective column density
$N_{\rm eff}$, for different density intervals and kinetic temperatures
$T_{\rm kin}$. This illustrates some basic features seen in the case of all
three model clouds.

$\Lambda/n$ decreases with $N_{\rm eff}$, particularly in the case of low
temperatures and high volume densities. The behavior is the result of the
dense cores becoming opaque. In low density regions, the net cooling is
reduced also by the incoming flux from surrounding regions that, due to the
higher density, have higher excitation temperature. At higher kinetic
temperatures more transitions are contributing to the cooling rates and photon
trapping cannot reduce the cooling rates to the same extent.

\begin{figure}[!th]
\epsscale{1.0}
\plotone{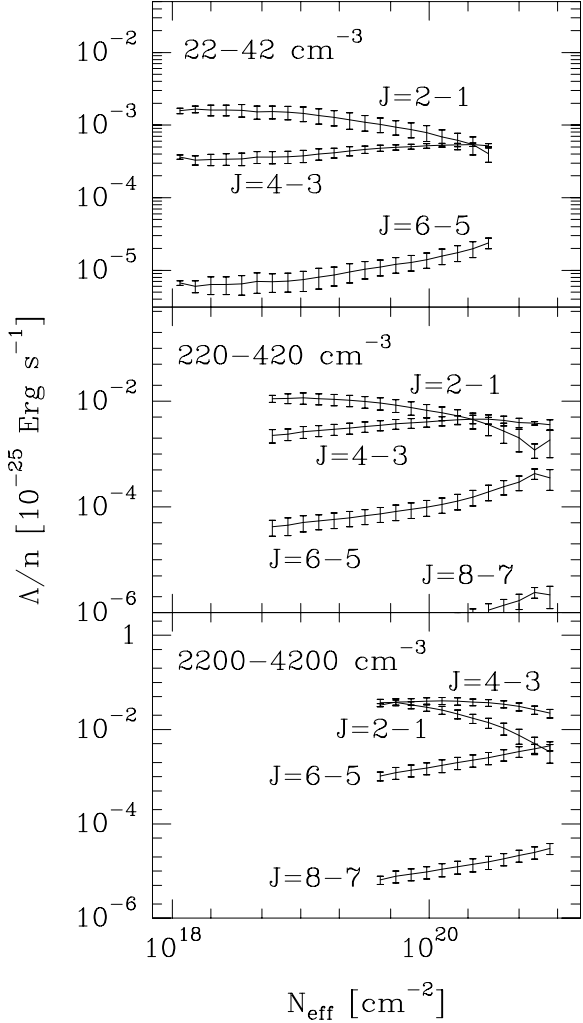}
\caption[]{%
Cooling rate $\Lambda({\rm CO})/n$ from different CO
transitions in model $B$, as a function of the effective
column density $N_{\rm eff}$. The kinetic temperature is $T_{\rm
kin}$=20\,K. The panels correspond to the density ranges:
2.2$\cdot$10$^1$-4.2$\cdot$10$^1$\,cm$^{-3}$ (a),
2.2$\cdot$10$^2$-4.2$\cdot$10$^2$\,cm$^{-3}$ (b) and
2.2$\cdot$10$^3$-4.2$\cdot$10$^3$\,cm$^{-3}$ (c). The variation
in the cooling rates within the given density intervals is
indicated by errorbars.}
\label{fig:coolcbt}
\end{figure}

Similar effects can be seen by studying the cooling rates $\Lambda/n$
computed separately for different CO transitions. Figure~\ref{fig:coolcbt}
illustrates the situation for model $B$. The increasing importance of 
higher CO transitions with increasing column density is evident.
In the higher density ranges the importance of higher transitions is larger
(see Figure~\ref{fig:coolcbt}). The increase in the cooling rate of the
$J=6-5$ transition, for example, tends to be more rapid than the decrease 
in lower transitions. As a result, the total cooling rate does not drop 
significantly with $N_{\rm eff}$.

\begin{figure}[!th]
\epsscale{1.0}
\plotone{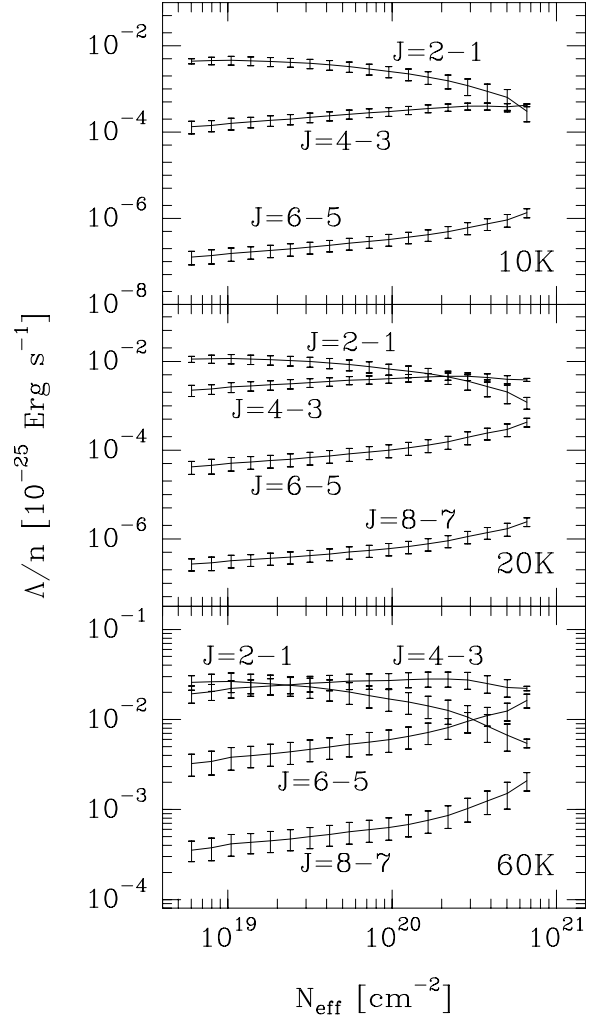}
\caption[]{%
Cooling rate $\Lambda({\rm CO})/n$ from different CO
transitions in model $B$, as a function of the effective
column density $N_{\rm eff}$ at three temperatures, $T_{\rm
kin}$=10\,K (top panel), 20\,K (middle panel) and 60\,K (bottom panel).
The curves represent the average rate for the density interval
2.2$\cdot$10$^2$-4.2$\cdot$10$^2$\,cm$^{-3}$.}
\label{fig:coolcbt_t}
\end{figure}

Figure~\ref{fig:coolcbt_t} shows similar $\Lambda/n$ dependencies at three
temperatures. At $T_{\rm kin}=10$\,K the cooling rate of the transition
$J=2-1$ (and $J=1-0$) decreases strongly as column density exceeds
$10^{20}$\,cm$^{-2}$. The increase in the cooling rate from the higher
transitions is unable to compensate for the loss and the total cooling rate
decreases as seen in Fig~\ref{fig:coolcolden}. On the other hand, already at
$T_{\rm kin}$=20\,K the cooling rate by the transition $J=4-3$ exceeds at
high column densities that of the transition $J=2-1$ and $\Lambda/n$ levels off.
However, the rates do not decrease significantly even at $N_{\rm eff}=5\cdot
10^{20}$\,cm$^{-2}$. At $T_{\rm kin}$=20\,K the average optical depths
averaged over the whole cloud are $\tau(J=2-1)\approx 130$, 
$\tau(J=4-3)\approx 65$ and $\tau(J=6-5)\approx 0.5$.
%% ##### @
At 60\,K the corresponding values are $\tau(J=2-1)\approx 30$,
$\tau(J=4-3)\approx 42$, $\tau(J=6-5)\approx 11$ and $\tau(J=8-7)\approx 0.4$.
At higher temperatures optical depths of individual transitions are lower and
this reduces the effect that column density has on local cooling rates.

\begin{figure}[!th]
\plotone{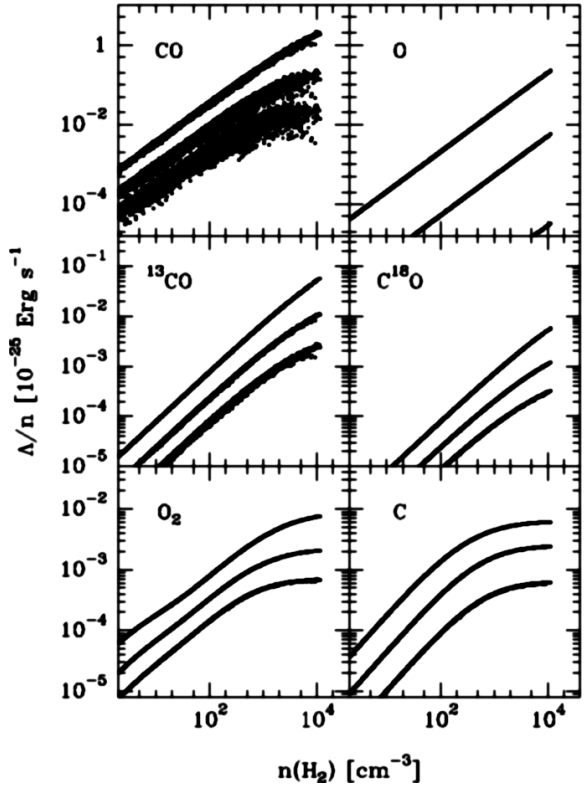}
\caption[]{%
The cooling rate $\Lambda/n$ in the cells of model $B$ as a function of local
gas density. The curves correspond to different values of the kinetic
temperature, $T_{\rm kin}$=10, 20, and 60\,K, in increasing order.
Rates are plotted for every tenth cell in the model. }
\label{fig:all_n_cool}
\end{figure}

\begin{figure}[!th]
\plotone{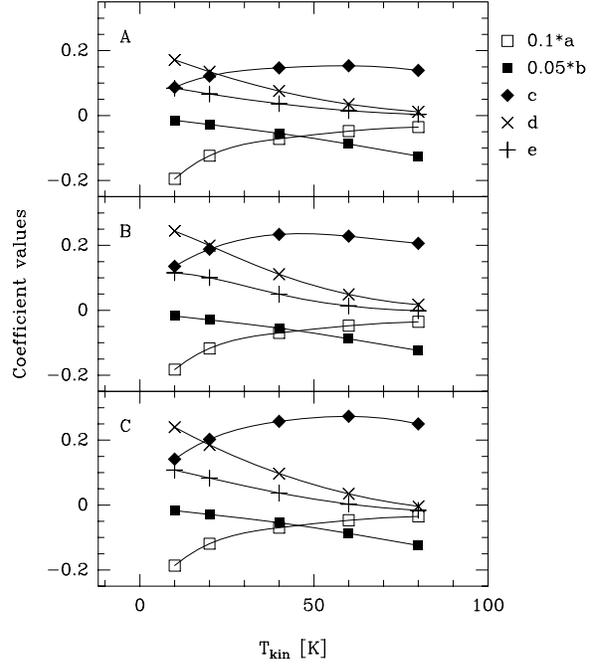}
\caption[]{%
Parameters of Eq.~\ref{eq:formula} as a function of the kinetic temperature of
the model, $T_{\rm kin}$, when fitted to the $^{12}$CO cooling rates
$\Lambda({\rm CO})/n$. The panels correspond to the three cloud models $A$,
$B$ and $C$.}
\label{fig:coeffs}
\end{figure}

\subsection{Analytic Approximations} \label{sect:analytical}

We have computed an analytic approximation of the cooling rates $\Lambda =
\Lambda(n, T_{\rm kin}, N_{\rm eff}$). The rates $\Lambda/n$ are fitted at
each temperature with a function
\begin{eqnarray}
f(n, N_{\rm eff}) & = &   a \times \nonumber \\
& & (1 + b\, log({n\over10^3}) + c\, (log({n\over10^3})^2 ))  
\times \nonumber \\
& & ( 1 + d\,log({N_{\rm eff}\over10^{20}}) + e\,(log({N_{\rm
eff}\over10^{20}}))^2 ) \nonumber \\
& = &   log\,\Lambda /n \,[10^{-25}\,{\rm erg}\,{\rm s}^{-1}]
\label{eq:formula}   
\end{eqnarray}
This is not the most optimal functional form for fitting the cooling rates but
is conceptually simple. Parameters $b$ and $d$ are, respectively, the slopes
of the density and the column density dependence and the parameters $c$ and
$e$ represent the non-linearity of these relationships. As a first
approximation the dependence of the $\Lambda/n$ on the density and column
density is linear on the log-log scale. Deviations from this behavior are
visible at low kinetic temperatures as a flattening or even a turnover in the
density dependence. Although similar turnover is seen in the cooling rates of
individual CO transitions also at higher temperatures (see
Figure~\ref{fig:all_n_cool}) the {\em total} cooling rates are monotonic in
the present density and column density ranges. The turnover is simply
transferred to higher densities and/or column densities. In the case of C{\rm
I}, flattening is even more pronounced and persists to higher temperatures.

For the fitting of Eq.~\ref{eq:formula} the cells of the clouds were
divided into small density and column density bins and the parameters of
Eq.~\ref{eq:formula} were fitted using the average values $\langle n\rangle$, 
$\langle N_{\rm eff}\rangle$ and $\langle \Lambda \rangle$ in each bin. 
The fitting was weighted with the number
of the cells in each bin and the fit is therefore least reliable in the tails of the
density and column density distributions.

Figure~\ref{fig:coeffs} shows the fitted parameters as a function of the
kinetic temperature in the case of $^{12}$CO. Note the decreasing dependence
of cooling rate on column density when kinetic temperature is
increased (coefficients $d$ and $e$). As already stated, this is a natural
consequence of the reduced optical depth per transition. A related effect is
the increased importance of density at higher temperatures (coefficients $b$
and $c$). 
% Depending on the local density the excitation temperature can lie
% anywhere between the temperature of the background radiation, $T_{\rm bg}$,
% and the kinetic temperature.
The first excitation level of CO is at $\sim$5\,K and e.g. $J=5$ is $\sim
80$\,K above the ground state. Therefore, at $T_{\rm kin}\sim$10\,K only a
couple of levels can be populated whereas at $T_{\rm kin}\sim$80\.K a density
increase can easily double the number of populated levels and markedly
increase the escape probability of the emitted photons.

\begin{figure}[!th]
\plotone{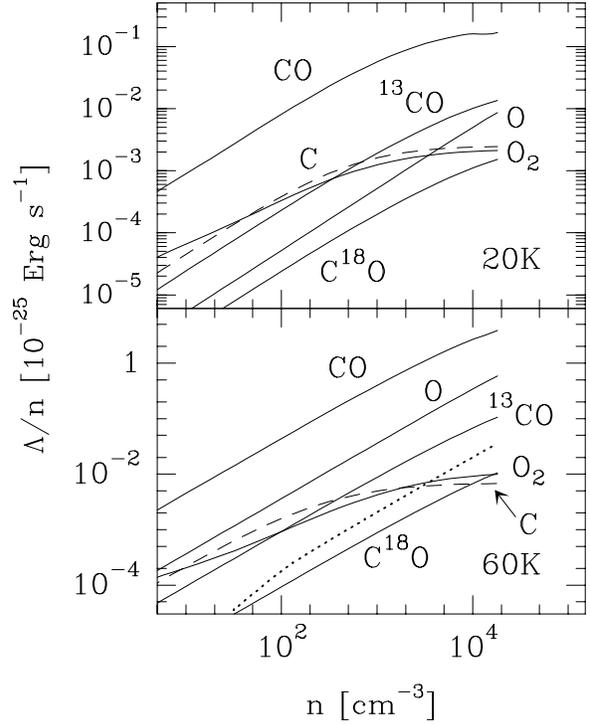}
\caption[]{%
Cooling rate, $\Lambda/n$, of the studied species as a function of the gas
density in model $B$. The lower panel corresponds to kinetic temperature
$T_{\rm kin}$=20\,K; the upper panel to $T_{\rm kin}$=60\,K. The rates for
C{\sc I} are drawn with dashed lines. At 60\,K the sum of the computed
ortho-H$_2$O and para-H$_2$O rates is shown as a dotted line. With the
abundances used in this paper the O$_2$ rates (not shown) would be roughly
equal with the C{\sc I} rates. As the real O$_2$ abundances are expected to be
much lower its contribution to the cooling will be negligible}
\label{fig:19o_relative}
\end{figure}

The dependence on the effective column density, $N_{\rm eff}$, is not very
strong for the cooling rate $\Lambda/n$. This is true for optically thin
species and even for $^{12}$CO, at least for temperatures above 10\,K (see
Figure~\ref{fig:coolcolden}). It is therefore reasonable to search for an
analytical approximation for $\Lambda/n$ as a function of density and
kinetic temperature alone. The function
\begin{equation}
\log{(\Lambda/n)} = c_1 \sqrt{\log{T}}   
+ ({\log{n}})^{3/2}    
\left(
c_2 + c_3/\sqrt{\log{T}}
\right)
\label{eq:ntfit}
\end{equation}
may be used to represent the $^{12}$CO and the total cooling rate with
$\sim$20\% accuracy over the studied density and temperature ranges, where
the absolute values of $\Lambda/n$ change by more than a factor of 10$^4$. The
parameter values obtained for the fits to total cooling rates are listed in
Table~\ref{table:ntfit}.

\begin{table}
\begin{center}
\caption[]{Parameters of Eq.~\ref{eq:ntfit} obtained by fitting
the cooling rate, $\Lambda/n$ [$10^{-25}$\,erg\,s$^{-1}$], in the three models.
The equation ~\ref{eq:ntfit} gives the local cooling rate as a function of
local gas density and the kinetic temperature. 
%% The cooling rate includes
%% cooling from $^{12}$CO, $^{13}$CO, C$^{18}$O, O$_2$, C{\sc I} and O{\sc I}
}
\label{table:ntfit}
\begin{tabular}{rrrrrrr}
\tableline
\tableline
model &   c$_1$ &   c$_2$ &   c$_3$ \\
\tableline
$A$   &   -4.23 &  0.80   &   -0.121 \\
$B$   &   -4.14 &  0.78   &   -0.118 \\
$C$   &   -4.24 &  0.82   &   -0.129 \\
\tableline
\end{tabular}
\end{center}
\end{table}

\section{Discussion} \label{sect:discussion}

In the parameter range studied $^{12}$CO is the main coolant. In
Figure~\ref{fig:19o_relative} we plot the cooling rates from model $B$ as a
function of the gas density, for all the coolants included in our study. The
most important difference between $T_{\rm kin}$=20\,K and 60\,K is the
increased cooling from O at the higher temperature, where it provides $\approx
10$\% of the total cooling rate. Due to the lower optical depth per transition
the $^{12}$CO cooling rate is, at 60\,K, considerably higher than the cooling
rate of $^{13}$CO and C$^{18}$O. The sum of the computed para-H$_2$O and
ortho-H$_2$O rates are shown for $T_{\rm kin}$=60\,K. At this temperature the
contribution from H$_2$O is not very significant.

According to \citet{neufeld95} the most important group of
coolants not included in our calculations is the non-hydride
molecules. Neufeld et al.\ estimate the total effect from these using the
formula
\begin{eqnarray}
L  = & \sum_{diatomic} L_{\rm CO}(0.01\,n, 100\,\tilde N) \nonumber \\
& + \sum_{polyatomic} L_{\rm CO}(0.008\,n, 7\,\tilde N)
\label{eq:neufeld}
\end{eqnarray}
with $L(\rm M)$ defined as $\Lambda = L\,n({\rm H}_2) n({\rm M})$ for each
species M. According to the formula the rates are similar to those of CO in
gas with lower volume density and higher column density. The average dipole
moment of diatomic non-hydrides (CS, NO, CN etc.) is approximately 10 times
the dipole moment of the CO molecule. The radiative rates are therefore two
orders of magnitude higher and this results in the first term of the equation.
The second term for polyatomic non-hydrides was based on the computed cooling
function of SO$_2$ (see \citealt{neufeld95}).

Using the CO cooling functions derived in this work, substituting $N_{\rm
eff}$ for $\tilde N$ in Eq.~\ref{eq:neufeld} and using the steady state
fractional abundances published by \citet{lee96}, `new standard model',
$T_{\rm kin}$=10\,K, $n$=10$^3$\,cm$^{-3}$, we can estimate the contribution
from these other molecules. With the CO cooling function fitted in model $B$
we find that at $T_{\rm}$=10\,K this amounts to $\sim$10\% of the CO cooling
rates at the high density limit of our models, $n$=$10^4$\,cm$^{-3}$. The
importance of these other molecules decreases rapidly with decreasing density.
The cooling rate decreases also with increasing kinetic temperature so that at
60\,K it is less than 5\% of the CO cooling rate.

Although the previous estimates are very crude we can conclude that for 
our models the non-hydride molecules are not important, except perhaps 
in the densest and coldest cores, where they could provide $\approx 10$\% 
of the total cooling. However, this number is uncertain and can be 
altered significantly, for example by assuming different fractional 
abundances.

\subsection{Comparison Between MHD Models} \label{sect:comparison}

Both density and velocity fields are important in determining the cooling rate
of optically thick lines. We may therefore expect to see some differences
between the MHD models although they are similarly inhomogeneous in both
density and velocity space.

Compared with the model $B$ at $T_{\rm kin}$=10\,K the average cooling rate,
$\lambda/n$, is $\sim$15\% lower in model $A$. The difference increases with
temperature and exceeds 30\% at $T_{\rm kin}$=60\,K. The difference is,
however, partly due to the fact that model $B$ has more high density cells
with correspondingly higher cooling rates (see Figure~\ref{fig:distributions}a). In
fact, for densities below 10$^3$\,cm$^{-3}$ the cooling efficiency is higher
in model $A$. At column density $N_{\rm eff}\approx$10$^{20}$\,cm$^{-2}$
and volume density 10$^2$\,cm$^{-3}$ the cooling rates ($\lambda/n$) in
model $A$ are a few of percent larger than in model $B$. At somewhat higher
densities, $n\approx3\cdot$10$^3$\,cm$^{-3}$, the rates are below those of
the model $B$ (at 10\,K by some 10\% but at 60\,K by only $\sim$1\%).

The net cooling rate ($\Lambda/n$) in model $C$ is 4\% smaller than in model
$B$. As a function of density and column density the rates are between those
of the models $A$ and $B$ and usually closer to the rates in model $A$. The
model $C$ included self-gravity which should affect the structure of the high
density concentrations. This does not, however, show clearly in the overall
density distribution (see Figure~\ref{fig:distributions}) and even at
$n\sim$10$^3$\,cm$^{-3}$ the cooling rates are quite similar to those found in
the other models. Same trend continues up to highest densities,
$n\sim$10$^4$\,cm$^{-3}$, with model $B$ having cooling rates 5-10\,\% in
excess of the other two models.

The cooling function should be sensitive to the cloud structure. In the
range $n=$10$^1$--10$^3$\,cm$^{-3}$ and $N=$10$^{19}$--10$^{21}$\,cm$^{-2}$
the differences in the $^{12}$CO cooling efficiency in the three MHD models
are, however, below 20\% and from this point of view the models do not differ
radically from each other. All have similar types of inhomogeneous density
structures and the overall velocity dispersions are also similar.
Furthermore, the clumpiness increases the photon escape probability and
reduces any effects caused by large optical depths. The differences between
MHD models would be enhanced e.g. in clouds with larger size, higher density
or lower turbulence.

\subsection{Average cooling rates} \label{sec:average}

In the previous chapters we have studied the local cooling rate as a
function of the {\em local gas properties}, density $n$ and kinetic
temperature $T_{\rm kin}$, and the general environment described by the
effective column density $N_{\rm eff}$. Since earlier theoretical studies and
observers tend to concentrate on the {\em average properties} of the clouds we
will now discuss the {\em average} cooling rate which is also proportional to
the net cooling of the cloud as a whole.

\begin{figure}[!th]
\plotone{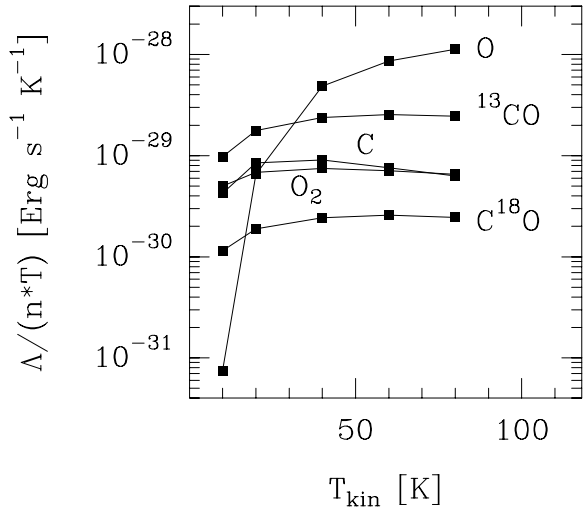}
\caption[]{%
The cooling rates averaged over model $B$, as a function of 
the kinetic temperature of the model. The mean density is 
$<n>$=320\,cm$^{-3}$.}
\label{fig:average_cooling}
\end{figure}

Figure~\ref{fig:average_cooling} shows the cooling rates in model $B$ averaged
over the model volume. Note the steep temperature dependence of the O{\sc I}
emission. The actual ratios between different species depend on the assumed
abundances and, as already mentioned, for species other than $^{12}$CO the
cooling efficiency scales almost linearly with the abundance.

We now study more closely the $^{12}$CO cooling. The cooling rate depends
strongly on the density (Figure \ref{fig:all_n_cool}). In an inhomogeneous
cloud this means that most of the cooling power comes from regions denser than
the average. The average cooling rate is also higher than the local rate at
the density equal to the mean density of the cloud. This is true as long as
the dense cores are not optically very thick. When the cores become opaque to most
of the cooling lines the cooling efficiency will not increase any more with
increasing density and the total cooling efficiency will be reduced.

\begin{figure}[!th]
\plotone{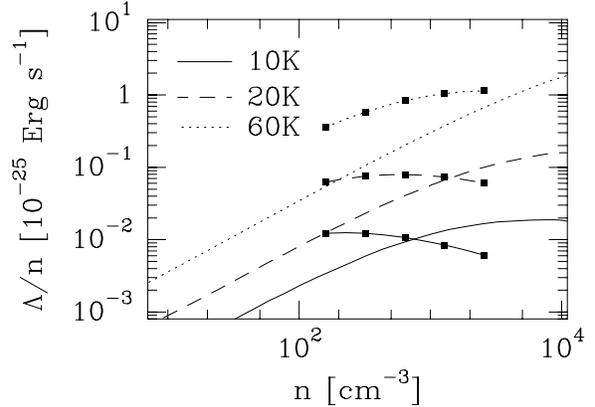}
\caption[]{%
The $^{12}$CO cooling in model $B$ at three kinetic temperatures. The longer
curves show the local cooling rates as a function of the local gas density.
The mean density of the model was 320\,cm$^{-3}$. The filled squares show the
cooling rates averaged over the model, as a function of the mean density of
the model. The models were obtained by scaling the mean density of model $B$
to 160, 320, 640, 1280 and 2560\,cm$^{-3}$}
\label{fig:ntnt}
\end{figure}

Figure~\ref{fig:ntnt} shows both the {\em local cooling rate} and the {\em
average cooling rate} of $^{12}$CO, as a function of density for three
different kinetic temperatures. The local rates are obtained from the
individual cells of the model $B$ as the cooling rate of a cell divided by the
gas density in the cell. The entire model cloud provides the point at
320\,cm$^{-3}$ for the curve showing the average cooling rate $<\Lambda>/<n>$.
Here the cooling rate and the density are averaged separately over the volume
of the model. The average cooling rate is proportional to the net cooling rate
of the entire model (erg\,s$^{-1}$). Other points for the average cooling rate
are from models which were obtained by scaling the mean density of model $B$
to 160, 640, 1280 and 2560\,cm$^{-3}$.

The rates follow the expected behavior outlined above. At lower densities the
average rates exceed the local rates, with a ratio approaching a factor of
ten. Even in the $<n>$=320\,cm$^{-3}$ model the CO rate $\Lambda/n$ levels off
at the highest densities. In models with high mean density this leads to a
turnover and eventually the average rate drops below the curve for the local
cooling rate. This means that gas below the mean density is getting more and
more important for the net cooling of the cloud. For 10\,K models the curves
meet below 10$^3$\,cm$^{-3}$. At higher temperatures the optical depth effects
are again reduced and at 60\,K the average rate remains above the local rate
until close to 10$^4$\,cm$^{-3}$.
% MIKA #11042001
This is perhaps the most striking result of the present work. 
%Molecular clouds
%can radiate ten times more efficiently then inferred from homogeneous models.
Molecular clouds described by turbulent density and velocity field as modeled
here, can radiate 10 times more efficiently than a uniform cloud with density
equal to the mean density of our model. As a consequence, their thermal
balance requires a ten times larger heating source.

In order to study separately the effects of the density and the velocity
inhomogeneities we have considered two further models that are based on
model $B$. The constant density model, $CD$, has constant density
320\,cm$^{-3}$ (equal to the mean density in the model $B$) but the velocity
structure of the model $B$. The constant velocity model, $CV$, has the same
density structure as the model $B$ but no macroscopic velocity field. Only
$^{12}$CO cooling rates were calculated since, due to high optical depth,
these are most likely to show differences between the models.

Compared with model $B$ the effective column densities are higher in $CV$ and
$CD$, since the inhomogeneity of either densities or velocities is removed.
This gives the appearance of increased $\Lambda/n$ when plotted against
$N_{\rm eff}$. This is, however, only due to the shift of the $N_{\rm eff}$
axis and the total cooling rate of the clouds is reduced. This is natural
since the escape probability of emitted photons is smaller.
% MIKA #11042001
For model $CV$ the total $^{12}$CO cooling rate is reduced by $\sim$10\% at
$T_{\rm kin}$=10\,K. The effect reduces at higher temperatures as the optical
depths of individual transitions are reduced. For model $CD$ the drop is
$\sim$50\% at 10\,K and increases with temperature. The difference between the
original model and $CD$ is not due to radiative transfer effects but is rather
a direct consequence of the steep density dependence of the cooling function.
In an inhomogeneous cloud most of the cooling is provided by regions with
density well above the average value.
For optically thin emission the ratio of average cooling rate and cooling rate
at the mean density can be derived directly based on the cell density
distribution (Figure~\ref{fig:distributions}) and the density dependence
$\Lambda \sim n^2$. The ratios are 3.9, 8.0 and 8.9 for the models $A$, $B$
and $C$, respectively.

%% ### @
Since density peaks are very important for the cooling we must make sure that
the results are not affected by the limited spatial resolution. The CO cooling
rates were compared in three variants of the model $B$ where the cloud was
divided into either 128$^3$, 90$^3$ or 48$^3$ cells. The plots of local
cooling rates against local density were almost indistinguishable although at
lower resolutions the total range of densities was reduced. A change in the
discretization causes changes in the cell densities. In the plot this means
only a small displacement along the $\Lambda(n)$ curve which itself remains
unchanged. As the cell size is increased some of the velocity dispersion
between cells (`macroturbulence') is transformed into turbulent velocity
inside the cells (`microturbulence') but this did not produce any noticeable
effects on the cooling rates. At $T_{\rm kin}$=20\,K the total cooling rates
of the models were within $\sim$3\% of each other. This shows that the
selected resolution was sufficient to capture the effects of density and
velocity inhomogeneities. Our models represent quiescent diffuse clouds. In
models with more developed cloud cores and steeper density gradients the
spatial resolution should be correspondingly higher.
%% ###

\subsection{Comparison with Earlier Work} \label{sect:others}

Most previous calculations of molecular line cooling rates have focused on
specific objects (for example dense cores) and are therefore not useful for
comparison with the present work. In particular, little has so far been
published on the subject in connection with inhomogeneous clouds. In the
following we look at some of the differences between our work and the results
obtained by \citet{goldsmith78} and \citet{neufeld95}. Their results apply to
clouds with continuous and smooth density distributions.
%%% ### @
Differences will be caused by two factors. Firstly, our MHD models contain a
range of densities and due to steep density dependence (e.g.
Figure~\ref{fig:all_n_cool}) total radiated energy will be larger than in a
homogeneous cloud with equal mean density (see previous chapter). Secondly,
excitation and escape probability of emitted photons will be affected by the
density and velocity inhomogeneities.
%%% ###

\begin{figure}[!th]
\plotone{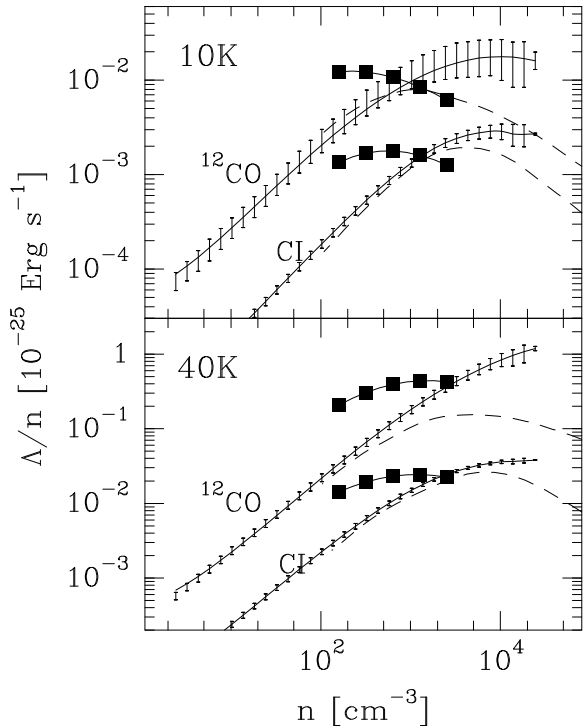}
\caption[]{%
Comparison of our $^{12}$CO and C{\sc I} cooling rates with
\citet{goldsmith78}. The solid curves show our local cooling rates in model $B$ as a
function of the local gas density. The errorbars indicate the variation within
the density bins. The solid squares show our estimates for the average cooling
as a function of the average density of the model cloud (see text). The
dashed curves are $^{12}$CO and C{\sc I} rates taken from
\citet{goldsmith78}}
\label{fig:ntfit}
\end{figure}

In Figure~\ref{fig:ntfit} we plot the \citet{goldsmith78} cooling rates for
$^{12}$CO and C{\sc I} together with our average rates at a few densities
(solid squares). In the same figure the local cooling rates for the
$n=320$\,cm$^{-3}$ model are also shown (solid curve). This should be compared
only with those points on the other curves that correspond to the same
density.

% We will first compare the CO rates. For the LVG model used by Goldsmith
% \& Langer the relevant parameter is the CO abundance divided by the velocity
% gradient, $X_{\rm CO}/\,(dV/dr)$. At 10\,K the curve shown in
% Figure~\ref{fig:ntfit} corresponds to a value of
% 1$\cdot10^{-4}$\,pc\,(km/s)$^{-1}$. With our abundance value,
% 5$\cdot$10$^{-5}$, this translates into a velocity gradient of
% 0.5\,km\,s$^{-1}$ corresponding to velocity shift $\sim$3\,km\,s$^{-1}$ over
% the size of out models, $L=6.25$\,pc. In our models the rms velocity
% dispersion over the whole cloud is $\sim$2\,km\,s$^{-1}$ at 10\,K. At 40\,K
% our rms turbulence is $\sim$4\,km\,s$^{-1}$ and we use Goldsmith \& Langer
% curve for $X_{\rm CO}/\,(dV/dr)$=4$\cdot$10$^{-5}$\,pc\,(km/s)$^{-1}$. This
% corresponds to turbulent width 7.8\,km\,s$^{-1}$. The average column density
% measures do, therefore, agree within a factor of 2. The situation is, however,
% more complicated in details, since in MHD models the velocity field is not
% completely random and density field is not homogeneous. There can never be a
% direct correspondence between MHD and LVG models. The MHD simulations are,
% however, closer to the situation in real clouds.

%{\bf

We will first compare the CO rates. For the LVG model used by Goldsmith
\& Langer the relevant parameter is the CO abundance divided by the velocity
gradient, $X_{\rm CO}/\,(dV/dr)$. In our models a corresponding parameter can
be calculated using the average velocity dispersion of the cloud and the
linear size of the model, $L$. Choosing models where these parameters agree
ensures that the average properties of the LVG and the MHD models will be
similar. Differences will be caused by differences at smaller scales i.e.
mostly by the density and velocity inhomogeneity of our models. In the MHD
simulations the velocity and density fields are not completely random and
depending on the line of sight the average density and column density can
differ significantly from the values averaged over the entire volume of the
cloud.

In the following we will use the FWHM of the one-dimensional velocity
distribution as the measure of the velocity dispersion in our models. The
value is calculated as the average over the whole cloud. In model $B$ the FWHM
of the velocity dispersion is 2.90\,km\,s$^{-1}$ at $T_{\rm kin}$=10\,K. With
abundance 5$\cdot$10$^{-5}$ and cloud size 6.25\,pc we get a column density
per velocity unit of 1.08$\cdot$10$^{-4}$\,pc\,km$^{-1}$s. In
Figure~\ref{fig:ntfit} the model is compared with Goldsmith \& Langer
calculations for $X/\,(dV/dr)$=10$^{-4}$\,pc\,km$^{-1}$s. At 40\,K the
velocity dispersion in the MHD model was higher, FWHM=5.77\,km\,s$^{-1}$,
corresponding to the higher speed of sound. The column density is
5.4$\cdot$10$^{-5}$\,pc\,km$^{-1}$s and the corresponding value of the
Goldsmith \& Langer model shown in Figure~\ref{fig:ntfit} is
4$\cdot$10$^{-5}$\,pc\,km$^{-1}$s. Goldsmith \& Langer note that their results
are not sensitive to the assumed velocity gradient and in our case similar
conclusion can be drawn from the weak column density dependence in
Figure~\ref{fig:coolcolden}. Therefore, an absolute equality of the column
density parameters is not crucial and the qualitative result of the comparison
would remain the same even if our column density values were scaled by a
factor of two.

At 10\,K our {\em local}\, $^{12}$CO cooling rate at $n=$320\,cm$^{-3}$ agrees
with the predictions of Goldsmith \& Langer (see Figure~\ref{fig:ntfit}). The
point is still on the linear portion of the curve i.e. radiative trapping is
not yet significant. The {\em average} cooling rate is, however, two times
higher than either the Goldsmith \& Langer value or our local cooling rate at
that density. This is due to the fact that the density dependence of the
cooling rate $\Lambda$ is steeper than $n^{1.0}$. In the case of a wider
distribution of density values the average cooling rate increases even when
the mean density remains unaltered.

At 40\,K the effects of optical depth are greatly reduced and in the model
with mean density 320\,cm$^{-3}$ the local cooling rate is almost a linear
function of local density. This increases the difference to Goldsmith \&
Langer results which are at $n=10^{4}$\,cm$^{-3}$ a factor of six below our
rate. At 10\,K the average cooling rate $\Lambda/n$ was a decreasing function
of density but at 40\,K it increases up to $<n>\sim10^3$cm$^{-3}$. In the
model with mean density $<n>$=320\,cm$^{-3}$ the local cooling rate exceeds
the volume averaged rate by a factor of four.

In Figure~\ref{fig:ntfit} the errorbars indicate the variation in local
cooling rate, $\Lambda/n$, within each density bin (see also
Figure~\ref{fig:all_n_cool}). Although this does include some noise from the
Monte Carlo calculations most of the variation is due to radiative transfer
effects. The cells are in different environments (e.g. cloud centre vs. cloud
surface) and this affects the cooling. At 40\,K the local cooling rate is
rather uniform while at 10\,K there is wider scatter, especially at high
densities. This shows again how an increasing kinetic temperature reduces the
effects of optical depth.

% }

% For C{\sc I} the abundance used in this paper was 10$^{-6}$. This is a factor
% of ten lower than the value used by Goldsmith \& Langer. In
% Figure~\ref{fig:ntfit} we plot their C{\sc I} rate for velocity gradient
% 1\,km\,s$^{-1}$\,pc$^{-1}$. At density 320\,cm$^{-3}$ this corresponds to a
% C{\sc I} column density 6.2$\cdot 10^{16}$\,cm$^{-2}$\,km$^{-1}$\,s. In our
% models column densities are 2.8$\cdot 10^{16}$\,cm$^{-2}$\,km$^{-1}$\,s and
% 1.4$\cdot 10^{16}$\,cm$^{-2}$\,km$^{-1}$\,s at 10\,K and 40\,K, respectively.
% Values were again calculated using the overall velocity dispersion. Despite
% the lower column density the average cooling power is roughly equal to the
% Goldsmith \& Langer values. At higher densities curves for the local cooling
% rate flatten and similarly the average rate depends only weakly on the mean
% density. Closer to $n=10^4$\,cm$^{-3}$ the ratio between Goldsmith \& Langer
% and our average rate approaches ten i.e. the ratio in the fractional
% abundances.

% {\bf 

For C{\sc I} the abundance used in this paper was 10$^{-6}$ i.e. a factor of
ten lower than the value used by Goldsmith \& Langer. The relevant parameter
is, however, again $X_{\rm CI}/\,(dV/dr)$. None of the results published by
Goldsmith \& Langer correspond exactly to the parameters of our models and
therefore we have rescaled our abundance value so that our model can be
compared directly with their Figure\,7. Another difference is caused by the
collisional coefficients. Goldsmith \& Langer used for C{\sc I}-H$_2$
collisions the \citet{launay77} C{\sc I}--H rates divided by ten. In this
paper we have used rates given by \citet{schroder91} and these result in a
significantly higher C{\sc I} cooling. However, in Figure~\ref{fig:ntfit} we
have derived the C{\sc I} cooling using the scaled \citet{launay77}
coefficients. After these modifications the predicted cooling rates agree at
low column densities and the mean density and the average column density per
velocity interval are identical to the values in the Goldsmith \& Langer
model.

The main features are similar as in the case of CO. Our local C{\sc I} rate
$\Lambda/n$ increases up to density $n\sim10^4$\,cm$^{-3}$ where the Goldsmith
\& Langer curves are already clearly decreasing. At 40\,K the turnover is
shifted further to a higher density. The average C{\sc I} rates at five mean
densities between 160\,cm$^{-3}$ and 2560\,cm$^{-3}$ are shown in the same
figure. In the model $<n>$=320\,cm$^{-3}$ the average rate exceeds the local
cooling rate at this density by a factor of $\sim$3. The ratio is the same at
both 10\,K and 40\,K. When the mean density exceeds 1000\,cm$^{-3}$ the
average rates drop close to Goldsmith \& Langer predictions.

Compared with CO the main difference is that the kinetic temperature has very
little effect on the shape of the C{\sc I} curves. At 40\,K the CO curve has
become almost linear while the C{\sc I} has similar flattening as at 10\,K.
The effect is visible also in the average rates. While for CO $\Lambda/n$
changes from a decreasing function to an increasing one, little change is seen
in the C{\sc I} curves. There are only three populated C{\sc I} levels and the
effect kinetic temperature can have on the optical depth of the transitions is
correspondingly smaller. The second excitation level of C{\sc I} is more than
60\,K above the ground state while CO has already five energy levels below
this.
% }

In comparison with \citet{neufeld95} some of the differences are due to the
abundances. For example, the chemical models of \citet{neufeld95} predict
O{\sc I} and O$_2$ fractional abundances slightly below 10$^4$. The values
adopted in this paper are lower by almost a factor of ten, but O{\sc I} still
provides a few per cent of the total cooling rate at the highest density and
temperature in our models.
The optical depth of most species is sufficiently low, so that the cooling
rate depends linearly on the abundance. For example, reducing the abundance of
O$_2$ by a factor of ten decreases the cooling rate by the same factor, the
accuracy of the linear relation being $\approx$3\% at 10\,K and better than
0.5\% at 60\,K. On the other hand, the $^{12}$CO cooling rate decreases with
$N_{\rm eff}$, especially at high column densities. Therefore, a change in the
$^{12}$CO abundance has a relatively small effect on the cooling efficiency.
We checked this by reducing the $^{12}$CO abundance by 50\% in the model $A$.
Assuming $T_{\rm kin}$=10\,K the value of $\Lambda/n$ is reduced by 47\% for
column densities $N_{\rm eff}\sim10^{19}$cm$^{-2}$ roughly independent of
volume density. However, when the column density reaches
$5\cdot10^{20}$cm$^{-2}$ the reduction in $\Lambda/n$ is no more than 20\%.

%%% ### @
\begin{figure}[!th]
\epsscale{1.0}
\plotone{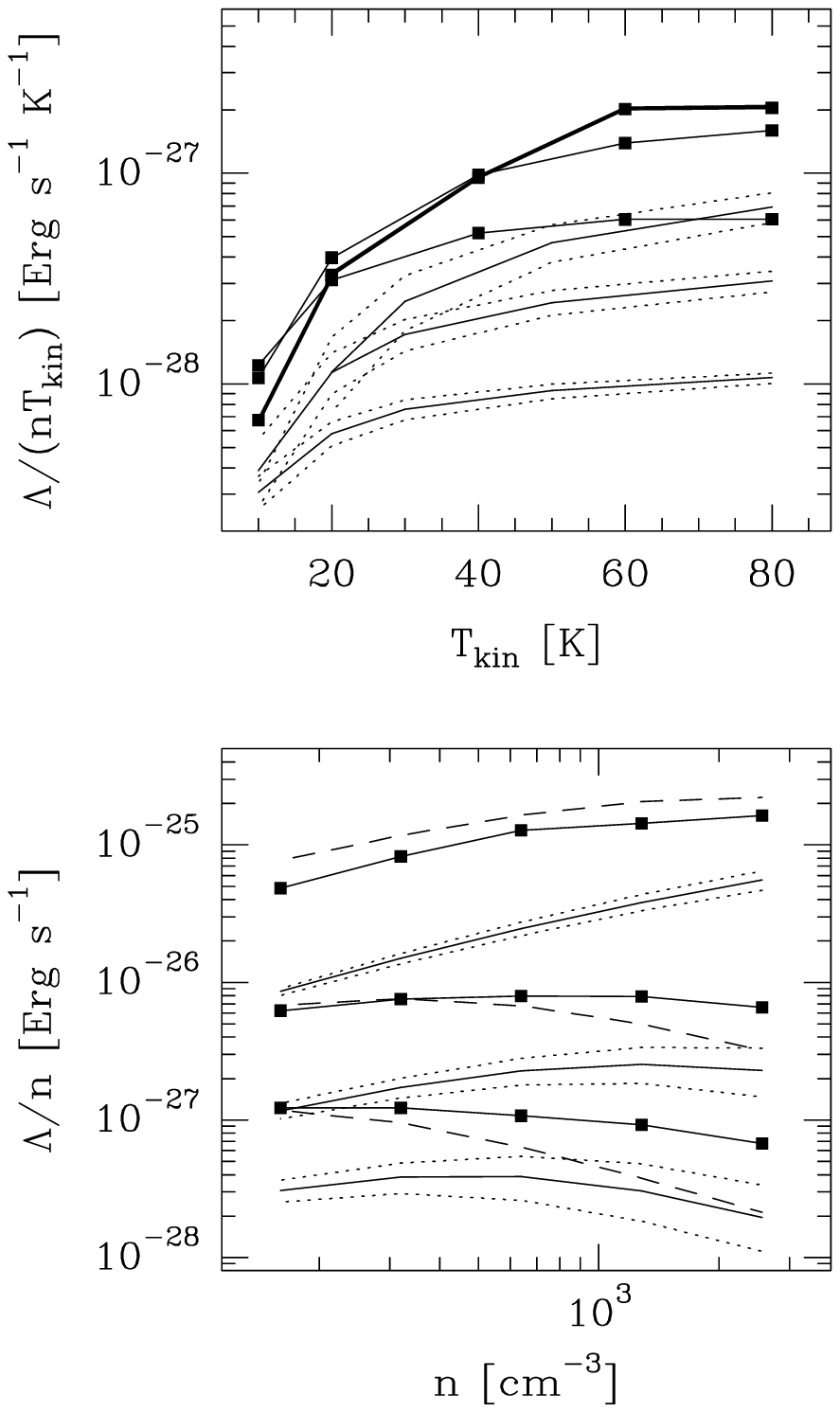}
\caption[]{%
Cooling rate of $^{12}$CO averaged over the volume of the model B. In the
upper panel the filled squares joined by lines are our results for cloud mean
densities $<n>$=160\,cm$^{-3}$ (lowest curve), 640\,cm$^{-3}$ and
2560\,cm$^{-3}$ (thick line). The corresponding rates computed from the
formulae presented by \citet{neufeld95} are shown with solid lines and dotted
lines correspond to column densities two times or half the estimated values of
$\tilde N$. In the lower panel the corresponding curves are plotted as the
function of density for temperatures 10, 20 and 80\,K. The dashed lines are
Neufeld et al. predictions for density 3$<n>$. }
\label{fig:average_cooling_co}
\end{figure}

\begin{figure}[!th]
\plotone{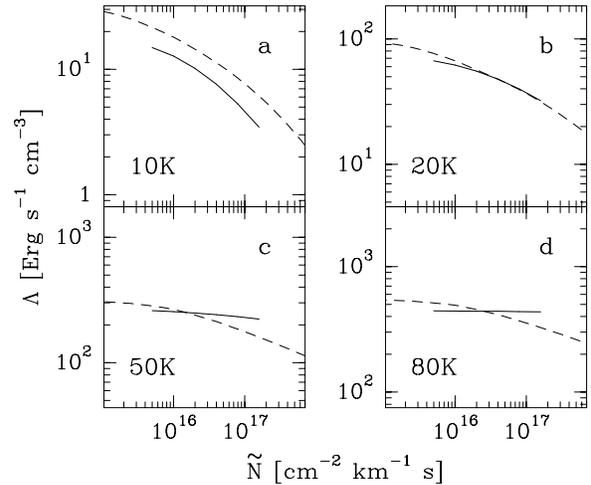}
\caption[]{%
Comparison of the cooling rates $\Lambda$ computed from the analytical
representation given by Neufeld et al. (\citeyear{neufeld95}; dotted lines) and
the rates in our model $A$ (solid curve). The curves correspond to density
$n\sim 10^3$\,cm$^{-3}$. The solid curve is computed from the analytical fit
of Equation~\ref{eq:formula} and is drawn only for the actual column density
interval present in the model.}
\label{fig:comparison}
\end{figure}

% {\bf 
\citet{neufeld95} give cooling rate as a function of column
density measure, $\tilde N = G n/|dv/dx|$, i.e. number density divided by the
velocity gradient or, in the absence of a gradient, by the velocity dispersion
divided by distance. This is essentially the number of hydrogen atoms per
velocity unit, in units of cm$^{-2}$\,km$^{-1}$\,s. There is, however, an
additional factor, $G$, which depends on the geometry. For plane parallel
medium with large velocity gradient the value is $G=1$ and in the centre of a
static sphere $G\sim0.5$. The effect of the abundance is included in the
parameter $\tilde N$ and its value need not be known separately.

In the upper panel of Figure~\ref{fig:average_cooling_co} we plot the {\em
average} $^{12}$CO rate for model $B$, with the mean density scaled to 160,
640, and 2560\,cm$^{-3}$. The parameter $\tilde N$ was estimated as
$G\,n\,L/\sigma$ with $G=1$ and $\sigma$ equal to the rms velocity dispersion
in the MHD model at given temperature. The lower three lines show predictions
by \citet{neufeld95} and the dotted lines show the rates with $\tilde N$
multiplied by two or by one half. The lower panel shows the corresponding
rates as function of the mean cloud density. The main difference is again due
to the density inhomogeneity of MHD models. Gas with density above the average
value provides most of the cooling power and the cooling rate is comparable to
that of a much denser homogeneous cloud. In the low density models the Neufeld
et al. formula would give a fairly good prediction of the cooling rate if
density value $n\sim3<n>$ were used. On the other hand, rate could be raised
only slightly by lowering $\tilde N$.

Next we will discuss again the {\em local} cooling rates. In
Figure~\ref{fig:comparison} our $^{12}$CO cooling efficiencies, $\Lambda$, are
shown together with the results of Neufeld et al. (\citeyear{neufeld95}, Table
3) as a function of $\tilde N$. Our curves correspond to the analytic
approximation of the local cooling rate in model $B$ (see
Sect.~\ref{sect:analytical}). For the plot we must determine a relation
between our parameter $N_{\rm eff}$, which is a quantity integrated over the
local absorption profile, and $\tilde N$ used by Neufeld et al. For gaussian
lines we have relation $N_{\rm eff}\approx$0.3\,$N/\sigma$ (see
Section~\ref{sect:rates}) while for plane-parallel flow (geometrical factor
$G=1$) $\tilde N = N/\sigma$. This gives a conversion $N_{\rm eff}\sim 0.3
\tilde N$ and this relation is used to plot our results on the $\tilde N$
scale in Figure~\ref{fig:comparison}. The scaling between $\tilde N$ and $N_{\rm
eff}$ is only approximate and depends on the LVG model assumed. The curves may
therefore be shifted in the horizontal direction. However, the shift should
correspond to no more than a factor of two change in  $\tilde N$.

The most important difference is the marked flattening of the column density
dependence that is seen in our models at higher kinetic temperatures. This is
the same effect as seen e.g. in Figures~\ref{fig:ntnt} and
\ref{fig:ntfit} and we interpret this as the consequence of varying excitation
conditions that allow efficient cooling even in the dense gas. Low density gas
radiates in low transitions while cooling in dense cores takes place mainly
through higher transitions. At higher kinetic temperatures more transitions
can contribute to the cooling and there is a clear difference between
subthermally excited low density gas and cores that are close to
thermalization. Different parts of the cloud are therefore decoupled not only
due to velocity differences but also because they radiate mostly in different
transitions. The difference between our results and those in \citet{neufeld95}
decreases with decreasing temperature, i.e., at low kinetic temperatures our
models behave more like homogeneous clouds. At low temperatures the number of
excited levels is small and the remaining transitions have higher optical
depth. In Figure~\ref{fig:comparison} we show curves only for density
$n\sim$10$^3$\,cm$^{-3}$ but the results are qualitatively similar even at
other densities.
%In this paper we have assumed for CO a fractional abundance
%value of $5\cdot 10^{-5}$\,cm$^{-3}$. The value used by \citet{neufeld95} need
%not be known since the comparison is based directly on the effective column
%density, $\tilde N$.

The detected differences between inhomogeneous and homogeneous cloud models
can be attributed to two effects. Firstly, density and velocity
inhomogeneities increase the photon escape probability. Both the excitation
and the escape probability of photons are determined by the optical depths
towards different lines of sight. In the center of a homogeneous cloud the
optical depth is the same towards all directions. An inhomogeneous cloud with
the same average optical depth has always a higher escape probability, since
the escape probability is proportional to the average of exp(-$\tau$) and not
to the average of $\tau$. This leads to higher escape probability but
generally also to lower excitation.
Secondly, in the considered models $\Lambda/n$ is still an increasing function
of density and therefore density variations tend to increase the total cooling
power. The slope of $\Lambda/n$ is, of course, determined by the radiative
transfer and the two effects are closely interrelated. The density
distribution was seen to affect the photon escape probability also indirectly
through excitation. Some excitation levels are populated only in the densest
regions and this leads to a partial decoupling between the dense cores and the
surrounding low density gas.

% }

\section{Summary and conclusions}

We have studied the radiative cooling of molecular gas at temperatures $T_{\rm
kin}$=10-80\,K and densities $n\la10^4$\,cm$^{-3}$, based on three-dimensional
MHD calculations of the density and velocity structure of interstellar clouds.
The models have been scaled to linear sizes $\sim$6\,pc and mean densities in
the range 160--2560\,cm$^{-3}$. The cooling rates for isothermal clouds were
computed by solving the radiative transfer problem with Monte Carlo methods.

We find that:
\begin{itemize}
\item
Inhomogeneous density and velocity fields reduce photon trapping and thus
increase the cooling rates. In comparison with homogeneous cloud models the
MHD models are much less affected by optical depths effects. This is
especially true at kinetic temperatures $T_{\rm kin}\ga$60\,K.
\item
There is a clear difference between the density dependence of local cooling
rates and the density dependence of cooling rates averaged over entire clouds.
\item
At low to intermediate densities most of the cooling power is provided by
clumps with densities above the average gas density. The average cooling rate
for a model with given mean density can be as much as an order of magnitude
larger than the local cooling rate at the same density. In models with higher
average density and lower temperature the differences are smaller since the
densest parts of clouds become optically thick.
\item
Compared with earlier models (\citet{goldsmith78}; \citet{neufeld95}) our
local cooling rates differ mainly at large densities where our rates are
higher due to reduced photon trapping. The volume averaged cooling rates are
higher than in the earlier models typically by a factor of few. At higher
temperatures ($T\ga 40$\,K) the difference can be almost one order of
magnitude.
\item
For the MHD models, the absence of the macroscopic velocity field would reduce
the cooling by up to 10\%.
\item
The absence of density fluctuations would reduce cooling by $\sim$50\% at
10\,K. This is caused mainly by the density dependence of the cooling rates
and the radiative transfer effects are less important. At high temperatures
($\ga$80\,K) the difference to homogeneous models approaches a factor of ten.
\item
$^{12}$CO is clearly the most important coolant over the whole parameter
range studied.
\item
At low temperatures $^{13}$CO is the the second most important coolant, after
$^{12}$CO. At temperatures $T_{\rm kin}>$60\,K it is exceeded by O{\sc I},
which can provide more than 10\% of the total cooling (assuming a relative
abundance $\sim$10$^{-5}$).
\item
In view of the recently observed very low O$_2$ and H$_2$O abundances these
species are unimportant for the cooling of the type of clouds studied in this
paper.
\end{itemize}

\begin{acknowledgements}
%% @
We are grateful to the anonymous referee for the useful comments received.
This work was supported by the Academy of Finland Grant no. 1011055.
\AA ke Nordlund acknowledges partial support by the Danish National Research
Foundation through its establishment of the Theoretical Astrophysics Center.
\end{acknowledgements}

\end{document}